\documentclass[showpacs,amsmath,amssymb,twocolumn]{revtex4}
\usepackage{graphicx}
\usepackage{dcolumn}
\usepackage{bm}

\begin{document}

\title{\bf\large{Nucleon Mass Splitting at Finite Isospin Chemical Potential}}

\author{\normalsize{Sheng Chang$^1$, Jifeng Liu$^1$, and Pengfei Zhuang$^2$}}

\address{$^1$College of Physics and Electronic Engineering, GuangXi
         Normal University, GuiLin 541004, China\\
         $^1$Physics Department, Tsinghua University, Beijing 100084, China}

\begin{abstract}
We investigate nucleon mass splitting at finite isospin chemical
potential in the frame of two flavor Nambu--Jona-Lasinio model. It
is analytically proved that, in the phase with explicit isospin
symmetry breaking the proton mass decreases and the neutron mass
increases linearly in the isospin chemical potential.
\end{abstract}


\maketitle

Recently, the study on QCD phase structure is extended to finite
isospin density. The physical motivation to study QCD at finite
isospin density and the corresponding pion superfluidity is
related to the investigation of compact stars, isospin asymmetric
nuclear matter and heavy ion collisions at intermediate energies.
While there is not yet precise lattice result at finite baryon
density, it is in principle no problem to do lattice simulation at
finite isospin density\cite{son}. It is found that\cite{kogut1}
the phase transition from normal phase to pion superfluidity
happens at a critical isospin chemical potential which is exactly
the pion mass in the vacuum, $\mu_I^c = m_\pi$. The QCD phase
structure at finite isospin density is also investigated in many
low energy effective models, such as chiral perturbation
theory\cite{son,kogut2,split,loewe,michae}, Nambu--Jona-Lasinio
(NJL) model\cite{toub,bard1,frank,he}, random matrix
method\cite{klein}, ladder QCD\cite{bard2}, and strong coupling
lattice QCD\cite{nishi}.

There are two kinds of isospin symmetry breaking. At low isospin
chemical potential, $\mu_I<\mu_I^c$, the isospin symmetry is
explicitly broken. In this phase, while $\sigma$ and $\pi$ mesons
are still the eigen collective modes of the Hamiltonian of the
system, the $\pi$ meson splits into three channels with linear
$\mu_I$ dependence for $\pi_+$ and $\pi_-$ masses\cite{he},
$M_{\pi_+}=m_\pi-\mu_I$ and $M_{\pi_-}=m_\pi+\mu_I$. At high
isospin chemical potential, $\mu_I>\mu_I^c$, the isospin symmetry
is spontaneously broken with nonzero pion condensate. In this
phase, $\sigma$ and $\pi$ are no longer the eigen collective modes
of the system, and the new eigen modes are the linear combination
of them. In the pion superfluid, the simple linear dependence of
the collective modes on the isospin chemical potential disappears,
and masses of the new eigen modes are determined by the
complicated pole equations\cite{he}.

Like the three $\pi$ mesons which form an isospin triplet, proton
and neutron are different isospin states of nucleon. Similar to
the linear $\mu_I$ dependence for charged pions, we guess that
there may exist a simple $\mu_I$ dependence for proton and neutron
masses in the phase with only explicit isospin symmetry breaking.
In this paper, we calculate the nucleon mass splitting at finite
isospin chemical potential in the NJL model. We first extend the
nucleon structure\cite{ishii} of a quark and a diquark in the
model to dense medium, and then determine analytically the proton
and neutron masses at finite baryon and isospin chemical
potentials.

We take the two flavor NJL model defined by the Lagrangian density
\begin{eqnarray}
\label{njl}
 \mathcal{L} & = & \overline{\psi}(i \gamma^\mu \partial_\mu
-m_0+\mu\gamma_0)\psi+G_s\left[\left(\overline{\psi}\psi\right)^{2}+\left(
\overline{\psi}i \gamma_5{\bf \tau}\psi\right)^2\right]\nonumber\\
&& +G_d\left[\overline{\psi}i\gamma_5
C\epsilon_f\epsilon_c\overline{\psi}^t\right]\left[\psi^t\left(iC^{-1}\gamma_5
\epsilon_f\epsilon_c\right)^t\psi\right],
\end{eqnarray}
where $C=i\gamma_2\gamma_0$ is the charge conjugation matrix, the
superscript $t$ denotes the transposition operator, $m_0$ is the
current quark mass, $\mu$ is the quark chemical potential matrix
in flavor space $\mu=diag\left(\mu_u,
\mu_d\right)=diag\left(\mu_B/3+\mu_I/2, \mu_B/3-\mu_I/2\right)$
with $\mu_B$ and $\mu_I$ being baryon chemical potential and
isospin chemical potential, $G_s$ and $G_d$ are coupling constants
in color singlet channel and anti-triplet channel, ${\bf
\tau}=\left(\tau_1,\tau_2,\tau_3\right)$ are Pauli matrices in
flavor space, and $\epsilon_f$ and $\epsilon_c$ are totally
antisymmetric tensors in flavor and color spaces.

Since we are interested in the nucleon mass splitting induced by
explicit isospin symmetry breaking at $\mu_I<\mu_I^c$, there are
no pion superfluid and color superconductor, the only condensate
is $\sigma=\left<\overline\psi\psi\right>$  which is the order
parameter of chiral phase transition.

In mean field approximation, the thermodynamics of the system can
be evaluated as\cite{he}
\begin{eqnarray}
\label{omega}
\Omega&=&{\left(m-m_0\right)^2\over 4G_s}\\
&&-3\int{d^3{\bf p}\over
(2\pi)^3}\sum_{i=u,d}\sum_{a=\pm}\left[E_i^a+T\ln\left(1+e^{-E_i^a/T}\right)\right]\nonumber
\end{eqnarray}
in terms of the quasi-particle energies $E_i^\pm=E_p\pm\mu_i$ with
$E_p=\sqrt{{\bf p}^2+m^2}$, where $m=m_0-2G_s\sigma$ is the
effective quark mass. From the thermodynamic stable condition of
the system, the physical condensate is determined by the gap
equation
\begin{equation}
\label{gap}
{\partial\Omega\over \partial\sigma}=0
\end{equation}
which gives the temperature and chemical potential dependence of
the quark mass, $m(T,\mu_B,\mu_I)$. At zero temperature, the quark
mass keeps its vacuum value in the normal phase without pion
superfluid and color superconductor\cite{he},
$m(0,\mu_B,\mu_I)=m(0,0,0)$.

Since we still have color symmetry, the effective quark propagator
$S$ in mean field approximation is independent of color index and
diagonal in flavor space,
\begin{equation}
\label{s}
S\left(p,\mu_B,\mu_I\right)=diag\left(S_u\left(p,\mu_u\right),S_d\left(p,\mu_d\right)\right)
\end{equation}
with the elements
\begin{eqnarray}
\label{sud} S^{-1}_i\left(p,\mu_i\right)&=&\gamma^\mu
p_\mu+\gamma^0\mu_i -m\nonumber\\
&=&S^{-1}_i\left(p_0+\mu_i,{\bf p},0\right),\ \ \ \ i=u,d.
\end{eqnarray}

In the NJL model, the meson modes and diquark modes are regarded
as quantum fluctuations above the mean field, they can be
calculated in the frame of RPA\cite{klevansky,zhuang,he}. When the
mean field quark propagator is diagonal in flavor and color
spaces, e.g. the case with only chiral condensate, the summation
of bubbles in RPA selects its specific channel by choosing at each
stage the same proper polarization function, the diquark
propagator can be written as
\begin{equation}
\label{diquark1} D_d\left(k,\mu_B,\mu_I\right)={4iG_d\over
1-2G_d\Pi_d\left(k,\mu_B,\mu_I\right)},
\end{equation}
where $\Pi_d$ is the diquark polarization function constructed by
two quarks\cite{ishii},
\begin{eqnarray}
\label{diquark2}
\Pi_d\left(k,\mu_B,\mu_I\right)&=&-i\int{d^4p\over(2\pi)^4}{\rm
Tr}\Big[\nonumber\\
&&\times\left(\gamma^5C\epsilon_f\epsilon_c\right)S^t\left(k-p,\mu_B,\mu_I\right)\nonumber\\
&&\times\left(C^{-1}\gamma^5\epsilon_f\epsilon_c\right)S\left(p,\mu_B,\mu_I\right)\Big].
\end{eqnarray}
There are three kinds of diquarks $d_{\overline 1}, d_{\overline
2}$ and $d_{\overline 3}$ constructed by quarks with colors $2$
and $3$, $1$ and $3$, and $1$ and $2$, respectively. Since the
quark propagator is color independent, they are degenerated.
Taking the trace in flavor and color spaces and employing the
relation
\begin{equation}
\label{ss}
CS(p,\mu_B,\mu_I)^tC^{-1}=S(-p,-\mu_B,-\mu_I),
\end{equation}
the polarization function can be simplified as
\begin{eqnarray}
\label{diquark3}
\Pi_d(k,\mu_B,\mu_I)&=&-3i\int{d^4 p\over
(2\pi)^4}{\rm
Tr}\big[\nonumber\\
&&\gamma_5S_d(p-k,-\mu_d)\gamma_5S_u(p,\mu_u)\nonumber\\
&&+\gamma_5S_u(p-k,-\mu_u)\gamma_5S_d(p,\mu_d)\big],
\end{eqnarray}
now the trace is taken only in Dirac space.

Taking the replacements $p_0\rightarrow p_0-\mu_u$ for the first
term in the square bracket and $p_0\rightarrow p_0-\mu_d$ for the
second term, the chemical potential dependence of $\Pi_d$ and
$D_d$ can be reflected in a shift of the diquark energy,
\begin{eqnarray}
\label{pid}
\Pi_d(k,\mu_B,\mu_I)&=&\Pi_d(k_0+{2\over 3}\mu_B,{\bf k},0,0),\nonumber\\
D_d(k,\mu_B,\mu_I) &=& D_d(k_0+{2\over 3}\mu_B,{\bf k},0,0).
\end{eqnarray}
It is easy to see that the diquark polarization and propagator are
isospin chemical potential independent. This is easy to be
understood. From the interaction in diquark channel in
(\ref{njl}), only the quarks with different flavor and color can
construct a diquark, and a $u$ quark and a $d$ quark have opposite
isospin chemical potential. We should emphasize two points for the
relation (\ref{pid}). First, (\ref{pid}) is valid only in normal
phase where the effective quark mass $m$ is a constant. In the
pion superfluid phase, $m$ is a function of chemical potentials,
$m(\mu_B,\mu_I)$, and then the relation (\ref{pid}) fails. Second,
at finite temperature the Lorentz invariance is broken down, and
the integration over quark energy $p_0$ or diquark energy $k_0$ is
changed into a summation over the discrete Fermi or Bose
frequencies. Therefore, the replacement used to get (\ref{pid})
does not work at finite temperature.

Now we investigate nucleons in the NJL model. A nucleon
constructed by three quarks can be described by the three-body
Faddeev equation\cite{ishii}. In the static
approximation\cite{buck}, the Faddeev equation is reduced to an
effective BS equation constructed by a quark and a diquark, the
quark propagator between two quark-diquark bubbles becomes a
constant $-1/m$, and then we can use RPA again to derive the
nucleon propagator\cite{mineo}
\begin{equation}
\label{nucleon1}
D_N(q,\mu_B,\mu_I)=\frac{\frac{3}{m}}{1+\frac{3}{m}\Pi_N(q,\mu_B,\mu_I)},
\end{equation}
where $\Pi_N$ is the nucleon polarization function
\begin{eqnarray}
\label{nucleon2}
\Pi_N(q,\mu_B,\mu_I)&=&-\int{d^4k\over(2\pi)^4}D_d(k,\mu_B,\mu_I)S_i(q-k,\mu_i)\nonumber\\
&=&-\int{d^4k\over(2\pi)^4}D_d(k_0+{2\over
3}\mu_B,{\bf k},0,0)\nonumber\\
&&\times S_i(q_0-k_0+\mu_i,{\bf q}-{\bf k},0).
\end{eqnarray}
Since the diquark contains a $u$ and a $d$ quark, we take
$S_i=S_u$ for protons and $S_i=S_d$ for neutrons. Making a
transformation for the diquark energy, $k_0\rightarrow
k_0-2\mu_B/3$, we have
\begin{eqnarray}
\label{nucleon3}
\Pi_N(q,\mu_B,\mu_I)&=&-\int{d^4k\over(2\pi)^4}D_d(k,0,0)\\
&&\times S_i(q_0-k_0+{2\over 3}\mu_B+\mu_I,{\bf q}-{\bf
k},0)\nonumber.
\end{eqnarray}
The chemical potential dependence of $\Pi_N$ and $D_N$ is only
reflected in the quark propagator.

From the definition of particle mass, nucleon mass $M_N$
corresponds to the pole of the nucleon propagator at zero
momentum,
\begin{equation}
\label{matter}
1+\frac{3}{m}\Pi_N(q,\mu_B,\mu_I)\Big|_{q_0=M_N,
{\bf q}=0}=0.
\end{equation}
In the vacuum it is reduced to
\begin{equation}
\label{vacuum}
1+\frac{3}{m}\Pi_N(q,0,0)\Big|_{q_0=m_N, {\bf
q}=0}=0
\end{equation}
for the vacuum mass $m_N$. At this stage, we can determine the
four parameters in the NJL model, the current quark mass $m_0$,
the momentum cutoff $\Lambda$ and the two coupling constants. By
fitting the pion mass $m_\pi=140$ MeV, pion decay constant
$f_\pi=93$ MeV and the quark mass $m=400$ MeV in the vacuum, we
have $m_0=5$ MeV, $\Lambda=593$ MeV and $G_s=6.92$ GeV$^{-2}$. The
coupling constant $G_d$ in the diquark channel is controlled by
the nucleon mass in the vacuum. The ratio between the two
couplings is $G_d/G_s=1.1$, corresponding to $m_N=940$ MeV and
diquark mass $M_d=600$ MeV which can de obtained from the pole of
the diquark propagator (\ref{diquark1}). Note that the value of
the ratio is in between the estimated minimum and maximum
values\cite{zhuang}.

From the comparison between the pole equations (\ref{matter}) in
matter and (\ref{vacuum}) in vacuum, we obtain analytically the
relation between the nucleon masses $M_N$ in matter and $m_N$ in
vacuum,
\begin{equation}
M_N+{2\over 3}\mu_B+\mu_i=m_N,\ \ \ \ i=u,d.
\end{equation}
For a proton which is a $u(ud)$ state and a neutron which is a
$d(ud)$ state, we have
\begin{eqnarray}
M_p&=&m_N-\mu_B-{1\over 2}\mu_I,\nonumber\\
M_n&=&m_N-\mu_B+{1\over 2}\mu_I.
\end{eqnarray}
Just as we expected, the nucleon mass splits into two channels in
matter at finite isospin chemical potentia $\mu_I<\mu_I^c$, and
the proton and neutron masses change linearly in $\mu_I$. Again,
we should note that these linear relations are true only at zero
temperature and in normal phase with only chiral condensate.

In summary, we have investigated the nucleon mass splitting in the
phase with explicit isospin symmetry breaking in the frame of two
flavor NJL model at zero temperature. While proton and neutron are
still degenerated at finite baryon chemical potential $\mu_B$ and
the mass drops down linearly in $\mu_B$, the introduction of
isospin chemical potential $\mu_I$ leads to their mass splitting.
The proton mass decreases and neutron mass increases linearly in
$\mu_I$. This remarkable mass splitting may significantly change
the properties, such as the threshold values, of those dynamic
processes including protons and neutrons in isospin asymmetric
nuclear matter.

At finite temperature, the pole equation for proton and neutron
masses is still valid but it needs numerical calculations to
obtain their temperature dependence. When we go into the pion
superfluid region with finite pion condensate, the nucleon mass
splitting is still true, but the linear relation fails, because
the quark propagator is off-diagonal in flavor space and the
diquark propagator and in turn the nucleon propagator become much
more complicated.

{\bf Acknowledgement:} One of the authors (S.C.) wishes to thank
the High Energy Nuclear Physics Group of Tsinghua University for
kind hospitality and helpful discussions with Lianyi He and Xuewen
Hao. The work was supported by the grants NSFC10135030, 10425810
and 10435080.


\end{document}